\def\aa{{A\&A}}
\def\aas{{A\&AS}}
\def\aj{{AJ}}
\def\annrev{{ARA\&A}}
\def\apj{{ApJ}}
\def\apjs{{ApJS}}
\def\baas{{BAAS}}
\def\mnras{{MNRAS}}
\def\nat{{Nature}}
\def\pasp{{PASP}}

\documentclass[cup5b]{caps}
\usepackage{graphicx}
\usepackage{amssymb}
\usepackage{ociwsymp3e}  

\HeadText{E. L. Blanton et al.}  

\begin{document}

\pagenumbering{arabic}

\author[]{E. L. BLANTON$^{1,2}$, M. D. GREGG$^{3}$, D. J. HELFAND$^{4}$,
R. H. BECKER$^{3}$, and R. L. WHITE$^{5}$
\\
(1) University of Virginia, Charlottesville, VA, USA\\
(2) Chandra Fellow\\
(3) University of California, Davis, CA, USA and IGPP/LLNL, Livermore, CA, USA\\
(4) Columbia Univsersity, New York, NY, USA\\
(5) Space Telescope Science Institute, Baltimore, MD, USA}

%
%

\chapter{Uncovering High-z Clusters Using \\ Wide-Angle Tailed Radio Sources}

\begin{abstract}

The morphologies of wide-angle tailed (WAT) radio sources (edge-darkened,
C-shaped, FR I radio sources) are the result of confinement and distortion
of the radio lobes by the dense X-ray-emitting gas in clusters or groups of
galaxies.  These radio sources are easily seen at high redshifts ($z\sim1$)
in short-exposure images from the Faint Images of the Radio Sky at 
Twenty-cm (FIRST) survey.  Using a sample of approximately 400 WAT sources
from the FIRST survey, we have discovered a number of high-$z$ clusters.
Here, we present the highest-$z$ cluster found so far using this method:
1137+3000 at $z=0.96$.  We include photometric and spectroscopic results.
Ten galaxies are 
confirmed at the cluster redshift, with a line-of-sight velocity dispersion
of $\sigma=530^{+190}_{-90}$ km s$^{-1}$, typical of an Abell richness
class 0 cluster. 

\end{abstract}

\section{Introduction}

Wide-angle tailed (WAT) radio sources were first defined as a class by 
Owen \& Rudnick (1976) who described them as a sub-class of head-tail
radio sources. Head-tail sources include a radio core and tails of
radio emission trailing from the core as the host galaxy moves through the
dense gas of a cluster's intracluster medium (ICM).  Head-tail galaxies also
include narrow-angle tails (NATs), which are generally associated with 
galaxies with large peculiar velocities in the outer parts of clusters; the
ram pressure from the interaction with the ICM bends the radio lobes to create
the NAT structure.  Owen \& Rudnick found that WATs, however, are most often 
associated with bright elliptical or cD galaxies in the cores of clusters, 
and surmised that their smaller peculiar velocities would therefore produce
smaller bends in the radio lobes, resulting in a wide-angle, rather than a
narrow-angle, tailed morphology.  Later work, by e.g. Eilek et al.\ (1984)
and O'Donoghue et al.\ (1993), found that the peculiar velocities of the host
galaxies of WATs were too small to produce the observed bending of the lobes
by ICM ram pressure.  In addition, O'Donoghue et al.\ (1993) further defined 
the class of WATs as radio sources usually fulfilling five criteria: (1) 
association with a dominant galaxy in a cluster center, (2) radius from the
core to the edge of each lobe $r > 50$ kpc, (3) a dramatic transition
from a collimated jet to a radio hot-spot
(4) large-scale bends of both radio lobes in a C-shape, (5)
radio power within a factor of three of the FR break.  Point (5) refers to
the break in radio power found by Fanaroff \& Riley (1974); FR I 
sources typically have radio powers $P_{1440} < 5 \times 10^{25}$ W/Hz, 
lobes that darken
at the edges and that are often distorted, while FR II sources are the classic
collinear doubles with $P_{1440} > 5 \times 10^{25}$ W/Hz and edge-brightened
lobes.  At low-$z$, FR I sources are usually found in clusters 
or groups of galaxies; FR II sources usually are not (Hill \& Lilly 1991).

Since a WAT's peculiar velocity relative to the cluster mean has been found
to be insufficient to distort the radio lobes into the observed angles from
ICM ram pressure, alternate scenarios have been proposed for the bending of
the lobes.  Eilek et al.\ (1984) suggested that the ICM may have large-scale
fluid flows, left over from the cluster's formation.
Burns et al.\ (1994) added that large flows in the ICM could
result from large scale cluster-cluster mergers.  The merging gas could have
velocities high enough ($\sim1000$ km/s) to produce ram-pressure
bending of the lobes, as shown in the simulations of Roettiger et al.\ (1993).  

\subsection{Sample}

WAT radio sources should trace out clusters of galaxies, and in 
particular, they may locate clusters that have recently merged.
Using a sample of 384 WAT and NAT radio sources selected from the first 3000
deg$^2$ of the VLA Faint Images of the Radio Sky
at Twenty-cm (FIRST) survey (Becker, White, \& Helfand 1995), we have 
confirmed the existence of dozens of 
clusters with redshifts up to $z=0.96$ (Blanton et al.\ 2000, 2001, 2003).  
Imaging and multi-slit mask spectroscopic follow-up on a moderate-to-high 
redshift subset of ten of these 
revealed eight clusters with redshifts as high as $z=0.85$ 
(Blanton et al.\ 2000).
A complete (area- and magnitude-limited) low-$z$ subset including 
40 of the 384 objects was observed
optically, and $\sim50\%$ of them were found to be in richness
class Abell 0 or greater clusters (Blanton et al.\ 2001).  The majority of 
the low-redshift sources that are found in clusters, based on our 
optical richness measurements, are detected in the ROSAT All Sky Survey 
(five out of six detected for $z<0.2$). Those that are not found in clusters, 
based on our 
richness measurements, are not detected in the RASS (only one out of nine 
detected for $z<0.2$).  In other words, the
combination of radio and optical data gives us a very clear indication for
the presence, or lack of, a cluster, as confirmed with X-ray observations.

In order to identify high-$z$ candidates, we obtained $R$-band images of 
fields
surrounding WATs that were blank to the limit of the Digitized 
Sky Survey (DSS).  Some of these had no optical identifications even down 
to the
limit of our images ($m_{R}\sim22-23$).  We then performed near-infrared (NIR)
observations at the positions of these sources in hopes of uncovering
very red, high-$z$ cluster galaxies.  

Here we present optical and NIR observations taken at the position of the 
WAT radio source, 1137+3000.  A combination of imaging and 
spectroscopy confirms that this source is located in a distant cluster of
galaxies with a redshift $z=0.96$.
We assume $H_{\circ} = 70$ km s$^{-1}$ Mpc$^{-1}$, $\Omega_M = 0.3$,
and $\Omega_{\Lambda} = 0.7$ throughout.

\begin{figure}
\centering
\includegraphics[width=7cm,angle=0]{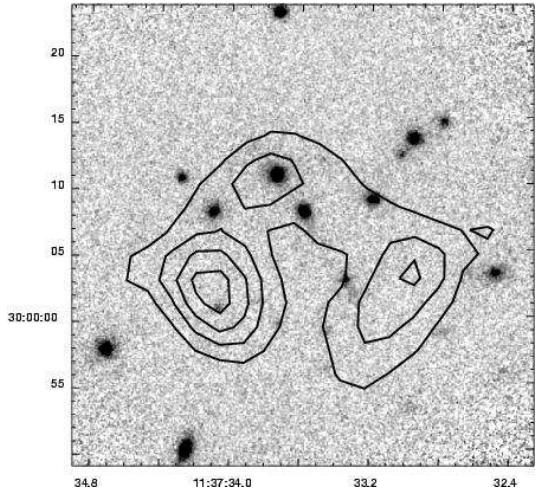}
\caption{Overlay of the {\it FIRST} 20cm radio contours onto a 
$K^{\prime}$-band image taken with the Keck NIRC.  The contours are
logarithmically spaced in the range 0.45 to 4.0 mJy.  With our assumed
cosmology, the figure is $280\times280$ kpc.\label{fig:ovly}}
\end{figure}

\section{1137+3000:  Radio Data}

The radio source is detected in the {\it FIRST} survey as a three component
object, including a core and two lobes.  
The source is clearly bent with an opening angle between the two lobes of 
$\sim 80^{\circ}$.  The size of the source (lobe to core to lobe)
is 30$^{\prime\prime}$
(240 kpc).  The total flux density, as measured by {\it FIRST} is 9.1 mJy.  
The flux density measured from the low-resolution
NRAO VLA Sky Survey ({\it NVSS}; Condon et al.\ 1997)
is 9.8 mJy, slightly higher than the {\it FIRST}
value. Using the {\it NVSS} flux density value, and assuming 
$P \propto \nu^{-\alpha}$ with $\alpha = 0.8$, the radio power of the 
source at 1440 MHz is $P_{1440} = 4.0 \times 10^{25}$ W Hz$^{-1}$.  
This power is slightly below the FR I/II break for radio galaxies 
(Fanaroff \& Riley 1974), and is typical of a WAT source (O'Donoghue et 
al.\ 1993).
An overlay of the {\it FIRST} radio contours onto a Keck near-infrared
camera (NIRC) image in the K$^{\prime}$ filter of the cluster
center is shown in Figure~\ref{fig:ovly}.  The figure is 280 kpc on a side
for our assumed cosmology.
The radio core is
coincident with an elliptical galaxy, as seen in the near-infrared image.
This NIR image, which covers only the inner 35$^{\prime\prime}$ region of the 
cluster, is used for display purposes only.


\begin{figure}
\centering
\includegraphics[width=7.5cm,angle=0]{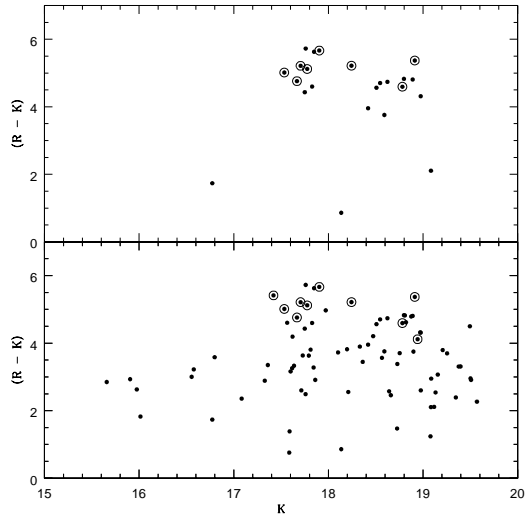}
\caption{$K$ vs. $(R-K)$ CMD.  The top panel includes 
the objects within a $1^{\prime}$ radius from the radio source; the bottom
panel includes the objects in the entire 15.6 arcmin$^2$ field.  There is
a concentration of objects with $4 < R-K < 6$ that are likely cluster
members. Spectroscopically confirmed cluster members are circled.\label{fig:krk}}
\end{figure}

\section{Color-Magnitude Diagrams}

The 1137+30 field was observed in the NIR with IRIM at the KPNO 2.1m on UT 
25 and 26 May 1997.  The total exposures
were 64 and 32 min. in the $K$-band and $J$-band, respectively.
The optical observations were obtained at the MDM 1.3m
with a total exposure in the $R$-band of 7200 s
on UT 21 May 1998 and were kindly provided by J. Kemp.

FOCAS was used to measure magnitudes for the standard stars and objects
in the target fields for the NIR and optical data.
Aperture magnitudes were measured for objects in the target fields. 
All images were scaled to have the same pixel scale ($0.^{\prime\prime}5$/pix).
The $R$ and $J$ images were smoothed
with a Gaussian to match the seeing in the $K$-band image (the worst of the 
three, with 1.$^{\prime\prime}$7).  We then used FOCAS to measure magnitudes
within an aperture with a diameter equal to 2$\times$FWHM
($3.^{\prime\prime}4$ = 6.8 pixels).  This is close to a total 
magnitude and will measure the same part of the galaxy in each of the bands.
At $z=0.96$ (see \S\ref{sec:spectra}), this aperture corresponds to 27 kpc
for our assumed cosmology.

The aperture magnitudes, measured as described above, for the radio source 
counterpart are 22.90$\pm0.04$, 19.46$\pm0.1$, and 17.78$\pm0.1$, in $R$, 
$J$, and $K$, respectively.
This gives colors of ($R-K$) = 5.12$\pm0.11$, ($J-K$) = 1.68$\pm0.14$, and 
($R-J$) = 3.44$\pm0.11$.
The absolute magnitude of the host galaxy is $M_{V} = -22.29$, assuming no
evolution, a redshift $z=0.96$ (see \S\ref{sec:spectra}), and a $K$-correction
from Coleman, Wu, \& Weedman (1980).  This is
typical of a bright elliptical galaxy, but fainter than a cD.

Color-magnitude diagrams (CMDs) were constructed for objects in the entire 
15.6 arcmin$^2$ field as well as for only the objects found within a 
$1^{\prime}$ radius aperture (480 kpc at $z = 0.96$, see \S\ref{sec:spectra}) 
from the radio source.
The $K$ vs. $(R-K)$ CMD is presented in 
Figure~\ref{fig:krk}.  The top panel includes the objects within 
$1^{\prime}$ from the radio source, while the bottom panel includes the 
objects from the entire field. The objects that are
confirmed to be at the cluster redshift (see \S\ref{sec:spectra}) are circled.
There is a much larger range in the colors of objects when looking at the 
CMD for the whole field.
The objects in the $1^{\prime}$ radius aperture are more concentrated on the
diagram than
those in the whole field, and are redder, consistent with membership in
a distant cluster.  Most of the objects within $1^{\prime}$ from the radio
source have colors in the range $4 < (R-K) < 6$, consistent with early-type
galaxies at $z\sim1$.

\section{Spectroscopy}\label{sec:spectra}

A long-slit observation (60 min.) of the cluster that included 
the radio counterpart and two
other galaxies was made at the Keck II 10m telescope with the Low Resolution
Imaging Spectrometer (LRIS; Oke et al.\ 1995) on 13 April
1998.
A multi-slit mask observation of the field surrounding the radio source
was made at the Keck II 10m with LRIS on 29 May 2001 with
a total exposure of 60 min.
Objects were chosen for the slit mask based on their $(R-K)$ colors.  Objects
that had $4 < (R-K) < 6$ were candidates for inclusion on the mask, since
they provide us the best opportunity of confirming galaxies at redshifts
$z\approx1$.  The mask included a total of twelve program object slits.

Spectra were successfully extracted for ten of the program objects.
Redshifts were initially 
estimated by identifying a few obvious lines, including the Ca II H+K break,
and the [O II] $\lambda3727$ and [O III] $\lambda\lambda 4959,5007$
emission lines.  The spectra were then Fourier cross-correlated with an
elliptical galaxy template using FXCOR in IRAF.

\begin{figure}
\centering
\includegraphics[width=7cm,angle=0]{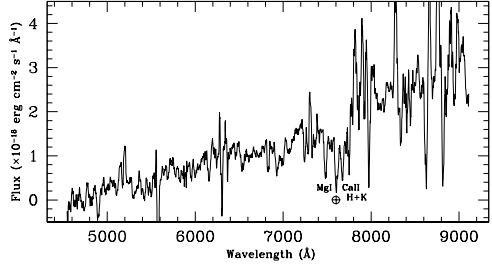}
\caption{Keck II LRIS spectrum of the radio source host galaxy.  The spectrum
has been smoothed with an 11 pixel boxcar.\label{fig:spect}}
\end{figure}

All ten program object spectra that were successfully extracted had redshifts in the range
$0.950 < z < 0.965$, giving us a $100\%$ success rate of identifying cluster
galaxies based on their NIR/optical colors.
The spectrum of the radio host
galaxy is shown in Figure \ref{fig:spect}, and is typical of an elliptical
galaxy.
We calculated the line-of-sight velocity dispersion using
$\sigma=\sqrt{(N-1)^{-1}\sum_{i=1}^{N}\Delta v_i^2}$, where
$\Delta v_i=c(z_{i}-\bar z)/(1+\bar z)$.
We find $\sigma=530^{+190}_{-90}$ km s$^{-1}$ ($68\%$ confidence).  
This is a typical value for an Abell richness 0 cluster 
(Lubin \& Bahcall 1993).

\section{Conclusions}

We have provided photometric and spectroscopic evidence at optical 
and near-infrared wavelengths for a high-redshift cluster of galaxies 
associated with a wide-angle tailed radio galaxy at $z = 0.96$.
The majority of high-redshift clusters found in traditional
X-ray and optical surveys select the most X-ray luminous and optically 
rich clusters.  The radio selection technique employed here is an important 
complement to those methods, in part because it can locate high-z clusters
with a wide range of X-ray luminosities and optical richnesses, thus aiding in
the study of galaxy evolution and cluster formation. 

\def\aa{{A\&A}}
\def\aas{{A\&AS}}
\def\aj{{AJ}}
\def\annrev{{ARA\&A}}
\def\apj{{ApJ}}
\def\apjs{{ApJS}}
\def\baas{{BAAS}}
\def\mnras{{MNRAS}}
\def\nat{{Nature}}
\def\pasp{{PASP}}

\end{document}